\documentclass{article}

\usepackage[english]{babel}

\usepackage[letterpaper,top=2cm,bottom=2cm,left=2.25cm,right=2.25cm,marginparwidth=1.75cm]{geometry}

\usepackage{graphicx}
\usepackage[colorlinks=true, allcolors=blue]{hyperref}
\usepackage[
    backend=biber,
    style=nature,
  ]{biblatex}

\usepackage{authblk}
\usepackage{pdflscape}
\usepackage{cleveref}
\usepackage{adjustbox,booktabs}
\usepackage{caption}

\def\fillandplacepagenumber{%
 \par\pagestyle{empty}%
 \vbox to 0pt{\vss}\vfill
 \vbox to 0pt{\baselineskip0pt
  \hbox to\linewidth{\hss}%
  \baselineskip\footskip
  \hbox to\linewidth{%
     \hfil\thepage\hfil}\vss}}
     
\usepackage{tikz,xcolor}
\definecolor{lime}{HTML}{A6CE39}
\DeclareRobustCommand{\orcidicon}{%
	\begin{tikzpicture}
	\draw[lime, fill=lime] (0,0) 
	circle [radius=0.16] 
	node[white] {{\fontfamily{qag}\selectfont \tiny ID}};
	\draw[white, fill=white] (-0.0625,0.095) 
	circle [radius=0.007];
	\end{tikzpicture}
	\hspace{-2mm}
}

\expandafter\xdef\csname orcidLC\endcsname{\noexpand\href{https://orcid.org/\csname orcidauthorLC\endcsname}{\noexpand\orcidicon}}

\title{Survey of Big Data sizes in 2021}
\author[1,2]{Luca Clissa \orcidLC{}}
\affil[1]{National Institute for Nuclear Physics, Bologna, Italy}
\affil[2]{University of Bologna, Department of Physics and Astronomy, Bologna, Italy}

\begin{document}
\maketitle

\begin{abstract}

The modern increase in data production is driven by multiple factors, and several stakeholders from various sectors contribute to it.
Although drawing a comparison of the sizes at stake for different big data players is hard due to the lack of official data, this report tries to reconstruct the yearly orders of magnitude generated by some of the most important organizations by mining several online sources.
The estimation is based on retrieving meaningful unitary data production measures for each of the big data sources considered, and the yearly amounts are then obtained by conjecturing reasonable per-unit sizes.
The 
final result is summarized in the form of a bubble plot.

\end{abstract}

\section{What are Big Data?}\label{sec:what}

In the last twenty years, we have witnessed an unprecedented and ever-increasing trend in data production. 
Hilbert and Lopez \cite{hilbert2011world} date the rise of this phenomenon back to 2002, with the beginning of the digital age.
Indeed, the transition from analog to digital storage devices enormously expanded the capacity of accumulating data, thus leading to the \emph{Big Data} era.

The term ``big data" was first introduced in 1990s \cite{16, 17} and it is commonly adopted to describe datasets whose size exceeds the potential to manipulate and analyze them within reasonable time limits \cite{snijders2012big}.
However, the expression does not target any specific storage size but rather assumes a deeper meaning that goes well beyond the sheer amount of data points.
In fact, big data embrace a broad spectrum of data sources including structured, semi-structured and, mostly, unstructured data \cite{dedic2016towards}.
Although multiple connotations have been attributed to the concept of big data over the years, a commonly shared definition is related to the so-called \emph{5 Vs} \cite{3}:

\begin{itemize}
    \item \textbf{Volume}: the actual quantity of generated data is huge, in the order of magnitude of terabytes and petabytes \cite{sagiroglu2013big}. More generally, it indicates amounts that are too large and complex to exploit conventional data storage and processing technologies;
    
    \item \textbf{Variety}: the data may come in several data types and from diverse origins. These include sources as sensors, social media, log files and more, plus they encompass heterogeneous formats like text, images, audio, video and so on; 
    
    \item \textbf{Velocity}: data are produced and/or processed at high rates \cite{kitchin2016makes}, typically nearly real-time;
    
    \item \textbf{Value}: data must carry valuable information that, if correctly analyzed, bring business value and profitable insights \cite{uddin2014seven}. In a scientific context, this means information that contribute to the advancement of human knowledge;
    
    \item \textbf{Veracity}: data sources must be reliable and generate high-quality data that can produce value \cite{onay2018review, 33};

\end{itemize}

Nonetheless, the community has not reached a complete agreement on the big data definition \cite{22, kitchin2016makes}, with some authors suggesting moving their characterization from the intrinsic properties to the techniques adopted to acquire, store, share and analyze the data \cite{balazka2020big}.

\section{How are Big Data produced?}\label{sec:how}

Besides the modification of the storage supply,  multiple factors significantly enhanced data production and, hence, favored the rise of the big data era.
In the first place, the diffusion of the internet and the progress of computer technologies provided more processing capabilities and easier access to data, thus stimulating further their production.
Consequently, several stakeholders as big tech companies, traditional industries, governments, healthcare institutions and more started increasingly contributing to this growth.
Finally, the introduction of \emph{smart} everyday objects that not only receive but also produce data exponentially accentuated individual contributions to the total data produced.
Modern objects, in fact, are endowed with technologies that allow to collect data and share them via a network -- the so-called Internet of Things \cite{ashton2009iot} --, thus augmenting the production rate even more.
For instance, sensors measuring the status and operation are now commonly used in industrial machinery and household appliances to ease their control and automate maintenance.
The same paradigm is also influencing the direction of the personal items market in various ways.
For example, some tech companies are recently investing in wearable devices like watches and glasses to enable the users to be always connected with a rapidly mutable environment, track their progress and explore the world in unparalleled manners thanks to virtual reality.
Furthermore, the solutions that digitization offers are being explored to respond to the emerging challenges of current times.
Think, for instance, of the urge for modernization of institutional processes posed by the pandemic. The massive spread of the infections has required unprecedented amounts of patients needing access to health assistance. However, the impossibility to scale up services and equipment correspondingly caused huge issues and jeopardized people's safety. In such context, the availability of intelligent systems capable of remotely monitoring patients' conditions and providing them with specialist support would have enormously helped.

By and large, the data production trends we are currently witnessing are driven by two main factors: the digital services offered by several stakeholders from different sectors and their use at scale by millions of users.
This work investigates this phenomenon by integrating multiple sources of various nature, aiming at providing reasonable and up-to-date ``\textit{guesstimates}" of yearly data production for some of the main big data companies.

\section{Big data sizes in 2021} \label{sec:sizes}

\begin{landscape}
\begin{figure}
    \centering
    \includegraphics[width=\linewidth]{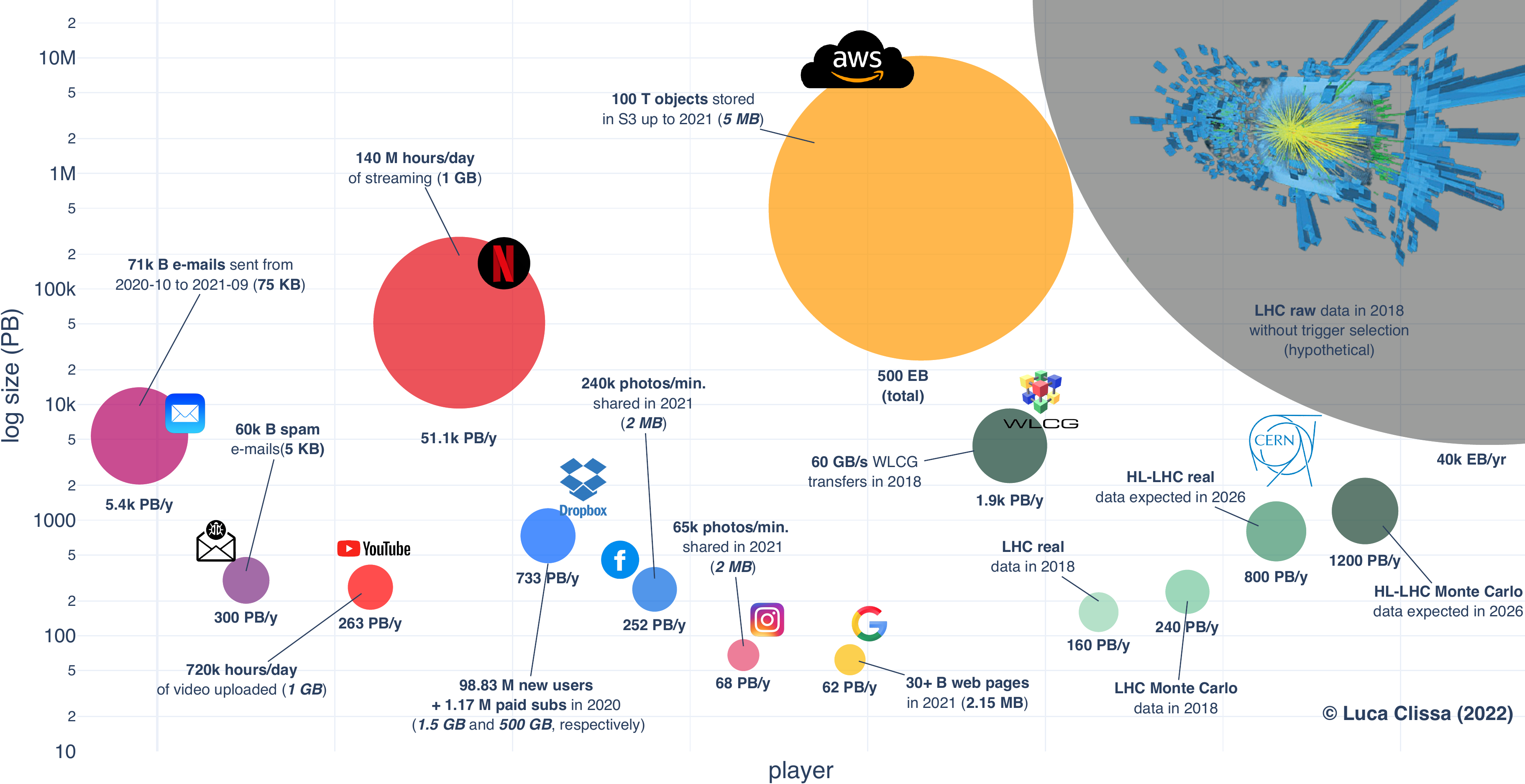}
    \caption{\textbf{Big Data sizes.} Bubble plot of the orders of magnitude of data produced by important big data players. The balloon areas illustrate the amount of data and the text annotations highlight the key factors considered in the estimates. Average per-unit sizes are reported in parentheses, where italic indicates measures reconstructed based on likely assumptions because no references were found. Interactive version available at: \href{https://clissa.github.io/BigData2021/BigData2021.html}{BigData2021.html}}
    \label{fig:bidata_size}
\end{figure}
\end{landscape}

The list of organizations that contribute to the creation and circulation of digital data in the modern society is nearly endless, including tech companies, media agencies, institutions, research centers and more.
Conducting an exhaustive survey for all of these stakeholders would be incredibly hard, if even possible.
For this reason, this work focuses only on a subset of the above entities and draws a comparison of their yearly data production.
In particular, miscellaneous online sources are extensively mined to retrieve information about the amount of contents produced, hosted or streamed by some of today's major players in the field of big data.
The corresponding yearly production rates are then obtained based on reasonable estimates of unitary sizes for such contents, e.g. average mail or picture size, average data traffic for 1 hour of video, and so on. 
In this process, considerations about storage space are omitted due to the lack of information regarding data management policies (e.g. data replication and redundancy). 
\Cref{fig:bidata_size} illustrates the results of such comparison, while \cref{tab:estimation-recap1,tab:estimation-recap2} summarize the estimation procedure and the sources of information considered.
However, the values reported are not meant to be point accurate, and they only give an idea of the orders of magnitude involved.

Despite not being the most popular among the mainstream audience, the CERN community \cite{cernwebsite}
is
one of the most prominent players concerning big data \textbf{production}.
Indeed, the physics experiments conducted by CERN scientists thanks to the Large Hadron Collider (LHC) \cite{lhcwebsite}
%
generated roughly 40k ExaBytes (EB) of raw data during its last run (2018) \cite{grandi2017HEPsize}.
To get an idea of how impressive this is, consider that over 100 trillion objects were reportedly stored in Amazon Simple Storage Service (S3) until 2021 according to Amazon Web Service (AWS) chief evangelist, Jeff Barr \cite{amazon2021objectscount}. 
Assuming a size of 5 MB per object in a representative average bucket (for example, see \cite{amazon2021objectssize}), this makes the total size of files ever stored in S3 equal to roughly 500 EB.
Thus, the data produced by LHC experiments in one year of operation is around one order of magnitude bigger than the total size of objects ever stored on Amazon cloud storage services.
%
%
However, storing such a tremendous amount of data is unattainable with current technology and budget. In addition, only a tiny fraction of that data is actually of interest, so there is no need to record all of that information.
Consequently, the vast majority of raw data is discarded straight away using hardware and software trigger selection systems, thus significantly lowering the amount of data \emph{recorded}.
As a result of this reduction, the actual acquisition rate
amounts to nearly 1 PetaBytes (PB) per day \cite{cern2017storage}, translating to roughly 160 PB%
\footnote{LHC registered 161 days of physics data taking in 2018 \cite{todd2018lhcAvail}}
a year in 2018. 
%
Besides the real data collected by LHC, physics analyses also require comparing experimental results with Monte Carlo data simulated according to current theories, thus producing somewhat between 1 and 2 times\footnote{a factor of 1.5 was adopted here for the bubble plot} additional data \cite{grandi2017HEPsize}.
Furthermore, the CERN community is already working at enhancing the Large Hadron Collider capabilities for the so-called High Luminosity LHC (HL-LHC) \cite{hllhc} upgrade.
As a result, the generated data are expected to increase of a factor $\geq5$ \cite{hllhc} and produce an estimated 800 PB of new data each year by 2026. 
Concerning other more renowned big data stakeholders as Google and Meta, the services they provide generate a yearly data production comparable to LHC effective figures, in the order of few hundreds of petabytes.
For instance, the Google search index tracked at least 30 billion webpages in 2021 \cite{van2016estimating, google2021index_size, djuraskovic2020googl_stats, indig2020index_size}, which gives a total of 62 PB considering an average page size of 2.15 MB \cite{http2021webpage_size}.
Regarding YouTube video uploads, instead, 720k hours of footages were uploaded daily \cite{dean2021youtube}, thus yielding something close to 263 PB if assuming an average size of 1 GB \cite{quora2021youtube}.
Similarly, the photos shared on Instagram and Facebook amount to an estimated 68 and 252 PB, respectively, given that 65k and 240k pictures where shared every minute on these social media \cite{domo2021infographic} and assuming 2 MB as the average picture size \cite{adobe2021fb_img_size}.
The yearly data production rises when considering storage services like Dropbox.
In 2020, the company declared 100 million new users, 1.17 millions of which were paid subscriptions \cite{dean2021dropbox}. By conjecturing that free accounts exploited 75\% of the 2 GB storage available, and that paid ones occupied 25\% of the total 2 TB, the amount of new storage required by Dropbox users in 2020 can be estimated to be around 768 PB.

Apart from the nominal values of the generated information, data \textbf{streaming} comprise a significant slice of the big data market.
As a matter of fact, the continual movement of small- to medium-sized files spawns massive traffic when scaled up to millions of users.
For instance, \textit{Statista} reports that nearly 131k billion electronic communications were exchanged from October 2020 to September 2021 (71k billion e-mails and 60k billion spam) \cite{statista2021mails}.
Assuming average sizes of \mbox{75 KB} and \mbox{5 KB} for standard \cite{lifewire2021avg_mail} and junk \cite{medium2014avg_spam} e-mails, respectively, that leads to an estimated 5.7k PB traffic in the analyzed period, which is clearly way beyond the amounts discussed so far.
Another example, is provided by Netflix, for which the scale is even larger.
The company's penetration, in fact, has sky-rocketed in the last years, also in response to the changed daily routines imposed by the pandemic. 
According to the 9\textit{-th} edition of the \textit{Data Never Sleeps} report by \textit{Domo}, 140 million hours of streaming per day were consumed by Netflix users in 2021 \cite{domo2021infographic}, which makes a total of roughly 51.1k PB assuming 1 GB of data for standard definition videos \cite{perry2021netflix}.
Perhaps surprisingly, the scientific community plays an important role also in this context.
Indeed, the LHC experiments are orchestrated by large collaborations formed by thousands of researchers spread around the world.
Hence, the data collected at CERN are continuously transferred thanks to the Worldwide LHC Computing Grid \cite{wlcgwebsite} to fuel innovative research. For example, 
a throughput of 60 GB/s was generated in 2018 \cite{wlcg2018throughput} thus giving a yearly projection of 1.9k PB, which is close to half of the global e-mails traffic and only one order of magnitude lower than Netflix usage.

\begin{table}
\ContinuedFloat*
\adjustbox{width=1.2\textwidth, center}{%
\begin{tabular}{@{}cccccccccc@{}}
\cmidrule(l){2-10}
 &
  YouTube &
  Dropbox &
  Facebook &
  Instagram &
  Google &
  \begin{tabular}[c]{@{}c@{}}LHC data\\ (real1)\end{tabular} &
  \begin{tabular}[c]{@{}c@{}}LHC data\\ (Monte Carlo)\end{tabular} &
  \begin{tabular}[c]{@{}c@{}}HL-LHC data\\ (real)\end{tabular} &
  \begin{tabular}[c]{@{}c@{}}HL-LHC data\\ (Monte Carlo)\end{tabular} \\ \midrule
\multicolumn{1}{c|}{\begin{tabular}[c]{@{}c@{}}production \\ unit\end{tabular}} &
  \begin{tabular}[c]{@{}c@{}}720k hours/day\\ video uploads\end{tabular} \cite{dean2021youtube} &
  \begin{tabular}[c]{@{}c@{}}100 M new users\\ (1.17 M paid subs) \cite{dean2021dropbox}\end{tabular} &
  \begin{tabular}[c]{@{}c@{}}240k photos/min\\ uploaded\end{tabular} \cite{domo2021infographic} &
  \begin{tabular}[c]{@{}c@{}}65k photos/min\\ uploaded\end{tabular} \cite{domo2021infographic} &
  \begin{tabular}[c]{@{}c@{}}30+ B\\webpages\cite{google2021index_size, djuraskovic2020googl_stats}\end{tabular} &
  \begin{tabular}[c]{@{}c@{}}161 days of\\ data taking \cite{todd2018lhcAvail}\end{tabular} &
  \begin{tabular}[c]{@{}c@{}}1.5 LHC\\ real data\end{tabular} &
  \begin{tabular}[c]{@{}c@{}}5+ times LHC\\ real data \cite{hllhc}\end{tabular} &
  \multicolumn{1}{c}{\begin{tabular}[c]{@{}c@{}}1.5 HL-LHC\\ real data\end{tabular}} 
  
  \\[5mm]

\multicolumn{1}{c|}{per-unit size} &
  1 GB \cite{quora2021youtube}&
  \begin{tabular}[c]{@{}c@{}}1 GB (free accounts)\\ and 400 GB (paid)\end{tabular} &
  2 MB \cite{adobe2021fb_img_size} &
  2 MB \cite{adobe2021fb_img_size} &
  2.15 MB \cite{http2021webpage_size} &
  1 PB \cite{cern2017storage} &
   &
   &
   
   \\[5mm]

\multicolumn{1}{c|}{period} &
  2021 &
  2020 &
  2021 &
  2021 &
  2021 &
  2018 &
  2018 &
  2026 &
  2026
  \\ \bottomrule
\end{tabular}%
}
\caption{\textbf{Summary of the estimation process.} The table reports a recap of the production units and the per-unit sizes considered in the estimation process for different big data sources, along with the corresponding time period}
\label{tab:estimation-recap1}
\end{table}
\begin{table}
\ContinuedFloat
\resizebox{\textwidth}{!}{%
\begin{tabular}{@{}ccccccc@{}}
\cmidrule(l){2-7}
 &
  Amazon S3 &
  LHC raw data &
  e-mails &
  spam &
  WLCG data trasfer &
  Netflix 
  
  \\ \midrule

\multicolumn{1}{c|}{\begin{tabular}[c]{@{}c@{}}production \\ unit\end{tabular}} &
  100 T objects \cite{amazon2021objectscount}&
  \multicolumn{1}{c|}{\begin{tabular}[c]{@{}c@{}}2400 M particle\\collisions per second \cite{grandi2017HEPsize}\end{tabular}} &
  \begin{tabular}[c]{@{}c@{}}71k B \\ mails sent \cite{statista2021mails}\end{tabular} &
  \begin{tabular}[c]{@{}c@{}}60k B junk \\ mails sent \cite{statista2021mails}\end{tabular} &
  throughput (s) &
  140 M hours/day \cite{domo2021infographic}
  
  \\[5mm]
  
\multicolumn{1}{c|}{per-unit size} &
  5 MB \cite{amazon2021objectssize}&
  \multicolumn{1}{c|}{1 MB \cite{grandi2017HEPsize}} &
  75 KB \cite{lifewire2021avg_mail}  &
  5 KB \cite{medium2014avg_spam} &
  60 GB \cite{wlcg2018throughput} &
  1 GB \cite{perry2021netflix}
  
  \\[4mm]

\multicolumn{1}{c|}{period} &
  up to 2021 &
  \multicolumn{1}{c|}{2018} &
  Oct 20 to Sep 21 &
  Oct 20 to Sep 21 &
  2018 &
  2021
  \\ \bottomrule

\end{tabular}%
}
\caption{\textbf{Summary of the estimation process (continued).}}
\label{tab:estimation-recap2}
\end{table}

\section{Conclusion}

The data production rate is at its peak, and it will continue growing in the next years. Conducting a punctual comparison of the amounts of information generated by the different organizations contributing to this trend is very hard, if even possible.
This work attempts to provide reasonable indications of the latest orders of magnitude of yearly data production for some of today's main players concerning big data.
Nevertheless, this study is not to be intended as a punctual estimation of the big data volumes produced by the individual organizations due to the lack of official sources. 
For the same reason, an important aspect as the amount of storage space occupied is omitted here since
more information about individual organizations' data management policies would be necessary.

One consideration that emerges from this survey is that streaming data already account for a significant slice of the big data market, and they will presumably continue to do so in coming years due to the increasing adoption of smart everyday objects able to generate and share data.

Another quite remarkable observation is that the experimental data collected by the scientific community play an important role in the big data phenomenon. Specifically, the amounts of data generated by nuclear physics experiments conducted at CERN are comparable to the traffic experienced by some of the most renowned commercial players as Google, Meta and Dropbox.

\printbibliography
\end{document}